\documentclass{PoS}

\title{Disconnected contributions to hadronic structure: a new method for stochastic noise reduction}

\ShortTitle{Disconnected contributions to hadronic structure}

\author{\speaker{Sara Collins}\\
        Institut f\"ur Theoretische Physik, Universit\"at Regensburg,\\
        93040 Regensburg, Germany\\
        E-mail: \email{sara.collins@physik.uni-regensburg.de}}

\author{Gunnar Bali\\
        Institut f\"ur Theoretische Physik, Universit\"at Regensburg,\\
        93040 Regensburg, Germany\\
        E-mail: \email{gunnar.bali@physik.uni-regensburg.de}}

\author{Andreas Sch\"afer\\
        Institut f\"ur Theoretische Physik, Universit\"at Regensburg,\\
        93040 Regensburg, Germany\\
        E-mail: \email{andreas.schaefer@physik.uni-regensburg.de}}

\abstract{We present a new method for reducing the stochastic noise of
  all-to-all propagators based on stopping the inversion of the
  propagator before convergence. The method is easy to implement,
  unbiased and independent of the quark action. Applying this method
  to the calculation of disconnected loops needed for hadronic
  structure observables we find savings in computer time of factors of
  $4-12$ depending on the operator inserted in the loop. When combined
  with a hopping parameter expansion technique we obtain combined
  gains of up to factors of $30$ for some operators.

}

\FullConference{The XXV International Symposium on Lattice Field Theory\\
                 July 30 - August 4 2007\\
                 Regensburg, Germany}
\newcommand{\dslash}{\! \not \!\! D}
\usepackage[cspex,bbgreekl]{mathbbol}

\begin{document}

\section{Introduction}
Nucleon structure observables such as baryon form factors and moments
of (generalised) parton distributions are extracted from 3pt functions
which have connected and disconnected contributions. The latter, of
the form $\mathrm{Tr}(\Gamma\mathrm{M}^{-1})\times$2pt function, are
normally omitted as they require the calculation of all-to-all
propagators $\mathrm{M}_{x,y}^{-1}$. Instead often differences between
observables, for example $g_A=\Delta u - \Delta d$, are quoted. However,
any settling of the question of the spin or the strangeness content of
the nucleon requires a calculation of the corresponding disconnected
loops. In the following we present a new method for calculating
all-to-all propagators which reduces the associated stochastic noise
and should make such calculations more viable. This is a general
method which can be applied to all cases where all-to-all propagators
are needed.
\subsection{Stochastic methods for all-to-all propagators}
\label{stochmethods}
The standard method for computing all-to-all propagators is via
stochastic sampling. A set of random complex $Z(2)$ noise vectors,
${|\eta^l\rangle}$, $l=1\ldots L$, is generated 
for which,
\begin{eqnarray}
\frac{1}{L}\sum_l| \eta^l\rangle\langle \eta^l| =  \mathbb{1} + \mathrm{O}(1/ \sqrt{L}),&\hspace{1cm}&
\frac{1}{L}\sum_l|\eta^l\rangle =   \mathrm{O}(1/ \sqrt{L}).\label{eqnnoise}
\end{eqnarray}
Using these vectors as sources one can construct an unbiased estimate
of the all-to-all propagator, $\mathrm{E}_L(\mathrm{M}^{-1})$, using
${|\eta^l\rangle}$ and the corresponding solution vectors
${|s^l\rangle}=\mathrm{M}^{-1}|\eta^l\rangle$:
\begin{eqnarray}
\mathrm{E}_L(\mathrm{M}^{-1}) =  \frac{1}{L}\sum_l|s^l\rangle \langle \eta^l|
              & = & \mathrm{M}^{-1} +  \mathrm{M}^{-1}\left(\frac{1}{L}\sum_l| \eta^l\rangle\langle \eta^l|-\mathbb{1}\right). \label{offdiag}
\end{eqnarray}
From eqns.~\ref{eqnnoise} and~\ref{offdiag} it is clear that the
stochastic error on the estimate only depends on the off-diagonal elements
of $\frac{1}{L}\sum_l| \eta^l\rangle\langle \eta^l|$ and falls off as $\mathrm{O}(1/
\sqrt{L})$. For a fixed number of configurations, depending on the
quantity studied (obtained using $\mathrm{E}_L(\mathrm{M}^{-1})$), the
stochastic noise can dominate over the gauge noise and additional
noise reduction techniques are required.

It is important to note that any noise reduction techniques should be
unbiased and the resulting reduction in noise should justify the
computational overhead. Existing techniques include
``partitioning''~\cite{Bernardson:1993yg} where each noise vector
$|\eta^l\rangle$ is replaced by a set of partitioned vectors
$|\eta^l\rangle_p$, $p=1\ldots P$, where $|\eta^l\rangle_p$ has many
zeros. By zero-ing entries in the source vector one will avoid some of
the large off-diagonal elements contributing in eqn.~\ref{offdiag};
the hope is that a smaller variance is obtained for the same amount of
computer time despite $P$ times as many inversions.
Wilcox~\cite{Bernardson:1993yg} found the gain (in terms of computer
time) for $\mathrm{Tr}(\Gamma \mathrm{M}^{-1})$ for colour-spin
partitioning to depend strongly on $\Gamma$ but could be in the region
of factors of $3-7$~(for $\Gamma=\gamma_{\mu\nu}$ and
$\gamma_\mu\gamma_5$) or higher (for $\gamma_5$).

The Kentucky group~\cite{hopex} take a different approach and use the
hopping parameter expansion~(HPE), to construct traceless estimates of
the off-diagonal elements in eqn.~\ref{offdiag}.  Subtracting these
estimates from $\mathrm{Tr}(\Gamma \mathrm{M}^{-1})$, where
$\mathrm{M}=1-\kappa\dslash$, leaves the trace unchanged but reduces the
variance. This approach should work well in the heavy quark regime,
for example for masses down to the strange quark mass.  Mathur and
Dong~\cite{mathurdong} found a gain of a factor of $6-7$ subtracting
up to $\kappa^4\dslash^4$ for the strangeness contribution to the magnetic
moment of the nucleon $G^s_M(0)$. The computational overhead of
performing the subtraction was not significant.

Additional approaches also exist: for example in the light quark
regime one can calculate the low lying eigenmodes of the Dirac
operator and use these to estimate part of the
propagator~\cite{eigen}. The remainder can be calculated
stochastically~\cite{stringbreak}. Different methods can often be
combined.

\subsection{A new approach: unbiased truncation of the solver}
\label{newapproach}
We present a new method for noise reduction which involves stopping
the inversion of the stochastic propagator before convergence, i.e.
using $n_t$ iterations in the solver to obtain
$|s^l_{n_t}\rangle=\mathrm{M}^{-1}_{n_t}|\eta^l\rangle$ instead
of running to convergence using $n_c$ iterations and obtaining
$|s^l_{n_c}\rangle=\mathrm{M}^{-1}_{n_c}|\eta^l\rangle$. The
difference between $\mathrm{M}^{-1}_{n_c}$ and $\mathrm{M}^{-1}_{n_t}$
can be estimated stochastically using an independent set of sources:
\begin{eqnarray}
\mathrm{E}[\mathrm{M}^{-1}_{n_c}]& = &
 \mathrm{E}_{L_1}[ \mathrm{M}^{-1}_{n_t}]+\mathrm{E}_{L_2}[\mathrm{M}^{-1}_{n_c}-\mathrm{M}^{-1}_{n_t}].\label{trunc}
\end{eqnarray}
This is based on an exact linear decomposition and the algorithm with
which both parts are calculated is well defined. Using two independent
sets of noise vectors for the two parts then implies an unbiased
estimate of $\mathrm{M}^{-1} _{n_c}$.  If the inverter converges
rapidly significant gains in computer time can be obtained. Rapid
convergence means that $\mathrm{M}^{-1}_{n_t}$ is very close to
$\mathrm{M}^{-1}_{n_c}$ even for small $n_t\ll n_c$.  Hence, the
stochastic error can be reduced by performing a large number of cheap
inversions for $\mathrm{E}_{L_1}[\mathrm{M}^{-1}_{n_t}]$, $L_1\gg L_2$,
and only a small number, $L_2$, of expensive inversions to calculate
the small correction.

To check this method we compared the exact result for
$(\mathrm{M}^{-1})_{x,y}^{s1c1,s2c2}$, where $s1c1$ denotes the spin
and color indices, $x=(0,0,0,3)$ and $y=(i,0,0,3)$, $i=0\ldots 10$, with an
estimate obtained from eqn.~\ref{trunc}. As expected we find
consistency within errors for different $n_t$, $L_1$ and
$L_2$.  For example, for $n_t=5$, $L_1=5500$, $L_2=300$, $i=1$, $s1=s2=1$,
$c1=c2=2$, $\mathrm{E}[\mathrm{M}^{-1}_{n_c}]=(0.0300(7),-0.0014(7))$
compared to the exact result of $(0.0302\ldots,-0.0010\ldots)$.

We now have two parameters, $n_t$ and the ratio $L_1/L_2$, which need
to be fixed, ideally, so as to minimize the variance of the
disconnected loop, $\mathrm{Tr}(\Gamma\mathrm{M}^{-1}_{n_c})$, at fixed
cost. For $L_1,L_2\gg 1$ the variance~(Var) is given by
\begin{eqnarray}
\mathrm{Var}_{L_1}[\mathrm{Tr}(\Gamma\mathrm{M}^{-1}_{n_t})]+\mathrm{Var}_{L_2}[\mathrm{Tr}(\Gamma(\mathrm{M}^{-1}_{n_c}-\mathrm{M}^{-1}_{n_t}))]&=& \frac{f_1}{L_1}+\frac{f_2}{L_2},
\end{eqnarray}
where $f_1$ and $f_2$ depend on $n_t$ and $\Gamma$, while the approximate
cost is given by
\begin{eqnarray}
C & = & L_1 n_t + L_2 n_c.\label{costtsm}
\end{eqnarray}
Using Lagrange multipliers and assuming $f_1$ to be approximately
independent of $n_t$ we obtain the optimal values
\begin{eqnarray}
n_t^{opt}  =  \frac{1}{n_c} \frac{f_2 f_1}{(f_2^\prime)^2}, & \hspace{1cm} &
\frac{L_1}{L_2}  =  \sqrt{\frac{f_1}{f_2}\frac{n_c}{n_t^{opt}} },\label{optvalues}
\end{eqnarray}
where $f_2^\prime=\partial f_2 / \partial n_t$. For our observables we find that using
these optimal values leads to $\frac{f_1}{L_1}\approx \frac{f_2}{L_2}$.

Additional gain can be obtained by combining with other noise
reduction techniques. Here we consider the HPE
approach~\cite{hopex,stringbreak}. The expansion of
$\mathrm{E}[\mathrm{Tr}(\Gamma\mathrm{M}^{-1}_{n_c})]$ to order $m$ is
given by:
\begin{eqnarray}
\mathrm{E}[\mathrm{Tr}(\Gamma\mathrm{M}^{-1}_{n_c})]&=&
\frac{1}{L}\sum_l \left[\langle \eta^l | \Gamma| \eta^l\rangle+ \langle \eta^l |\Gamma\kappa\dslash| \eta^l\rangle+
\ldots+\langle \eta^l |\Gamma\kappa^{m}\dslash^{m}| \eta^l\rangle\right]\nonumber\\
& & +\mathrm{E}[\mathrm{Tr}(\Gamma\kappa^{m+1}\dslash^{m+1}\mathrm{M}^{-1}_{n_c})],
\end{eqnarray}
where, since this is a geometric series, the last term gives the
remainder, $\sum_{p=m+1}^{\infty}\langle \eta^l |\Gamma\kappa^{p}\dslash^{p}| \eta^l\rangle$, averaged
over stochastic sources. One can omit terms in the expansion which
only contribute to the noise. All odd terms,
$\mathrm{Tr}(\Gamma\dslash^{2m+1})=0$, $\forall$ $\Gamma$. For the even terms,
$\mathrm{Tr}(\Gamma)=0$ $\forall$ $\Gamma\neq \mathbb{1}$, while~\footnote{These terms
  are zero for the Wilson action. For the clover action only the $m=0$
  term can be omitted.}  $\mathrm{Tr}(\Gamma\dslash^2)=0$ $\forall$ $\Gamma$ and for
$\Gamma=\gamma_\mu\gamma_5$ and $\gamma_5$, even
$\mathrm{Tr}(\Gamma\dslash^4)=\mathrm{Tr}(\Gamma\dslash^6)=0$.  Hence, for
$\Gamma=\gamma_\mu\gamma_5$ and $\gamma_5$, since all terms up to 8th order only contribute
to the noise, an improved estimate of the trace is given by
$\mathrm{E}[\mathrm{Tr}(\Gamma\kappa^{8}\dslash^{8}\mathrm{M}^{-1})]$.  For all
other $\gamma$ combinations we use~\footnote{Where for $\Gamma=\mathbb{1}$ we
  construct the non-vanishing $\mathrm{Tr}$ $\mathbb{1}$.}
$\mathrm{E}[\mathrm{Tr}(\Gamma\kappa^{4}\dslash^{4}\mathrm{M}^{-1})]$.  The 4th
and 6th order terms can be calculated explicitly~\cite{hopex} to
achieve the same level of improvement as for $\Gamma=\gamma_\mu\gamma_5$ and $\gamma_5$,
however, we have not done so in this study. To combine with our
truncated solver method we substitute, for example,
$\kappa^{8}\dslash^{8}\mathrm{M}^{-1}$ for $\mathrm{M}^{-1}$ in
eqn.~\ref{trunc}.
\section{Results}
We have performed an exploratory study of our method using
configurations provided by the Wuppertal group: these are $n_f=2+1$
dynamical configurations generated using a Symanzik improved gauge
action and a stout-link improved staggered fermion action. The lattice
spacing is fairly coarse, $a^{-1}\approx 1.55$~GeV while the volume is
around $2$~fm. Further details can be found in~\cite{latdetails}. For
valence quarks we used the Wilson action with $\kappa=0.166$, $0.1675$
and $0.1684$ corresponding to pseudoscalar masses of about $600$,
$450$ and $300$~MeV respectively. Our main results were obtained using
the conjugate gradient algorithm with even-odd preconditioning to
perform the propagator inversions.  However, section~\ref{bicgstab}
will show results obtained using the stabilised biconjugate gradient
algorithm (BiCGStab). The code used throughout was a modified
version of the Chroma code~\cite{chroma}.

Results are presented below for the disconnected loop, $\mathrm{Tr}(\Gamma
\mathrm{M}^{-1})$, where we have considered $\Gamma=\mathbb{1}$, $\gamma_\mu$, $\gamma_\mu\gamma_5$,
$\sigma_{\mu\nu}$, $\gamma_5$. Using $\mathrm{M}^{-1}=\gamma_5 (\mathrm{M}^{-1})^\dagger\gamma_5$ one can show that
the trace is either real or imaginary~\footnote{Of course the
  path integral expectation value $\langle \mathrm{Tr}\Gamma \mathrm{M}^{-1}\rangle=0$
  $\forall$ $\Gamma\neq \mathbb{1}$.}.  At this initial stage we are only interested in
the stochastic error and, hence, the results are presented for the
trace on a single configuration. In addition to combining our method
with the HPE approach we also partition in time: $|\eta^l\rangle$ are only
non-zero for $t=3$. 
\subsection{Truncating the solver}

\begin{figure}
\centerline{
\includegraphics[width=.4\textwidth]{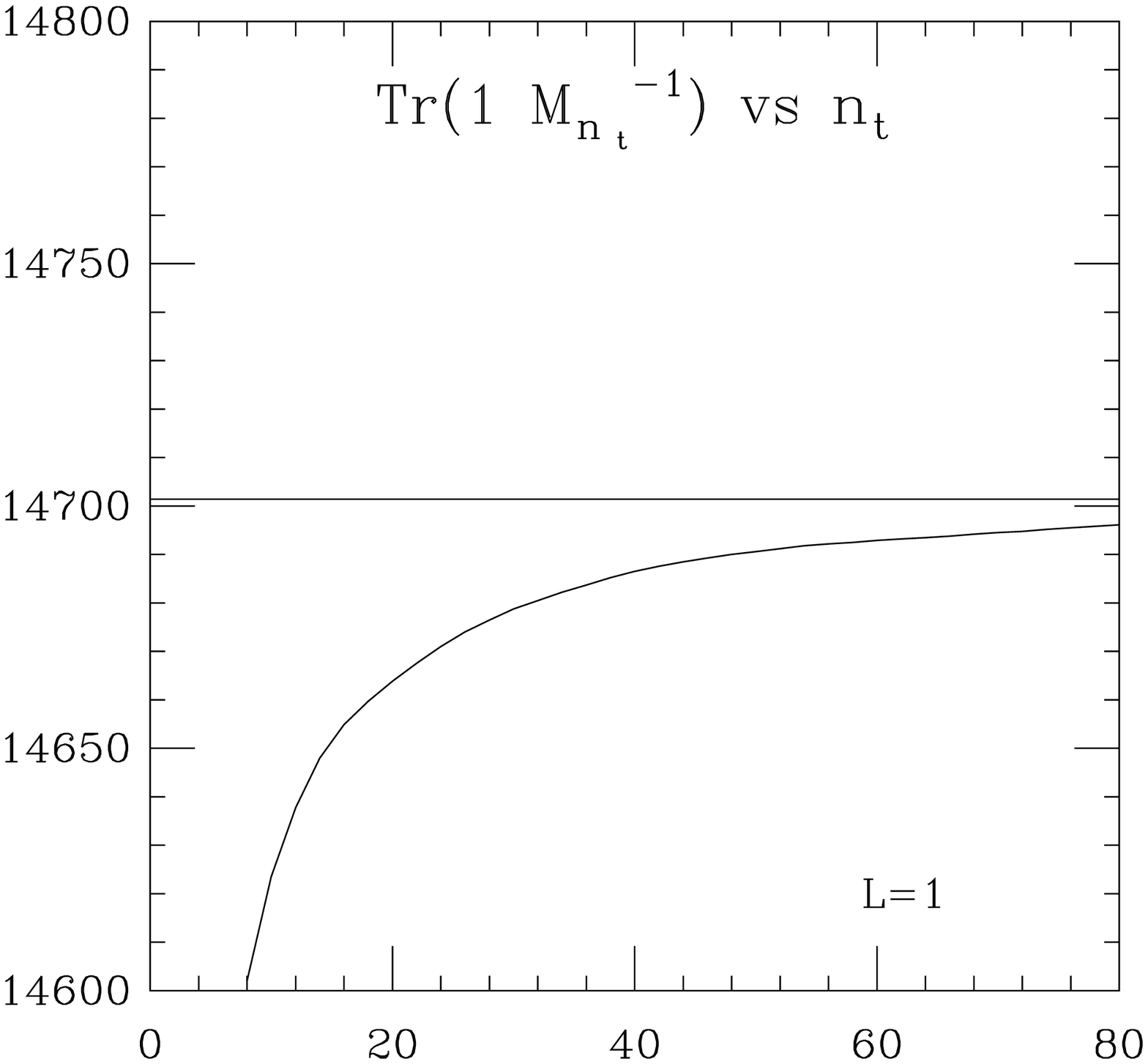}
\includegraphics[width=.4\textwidth]{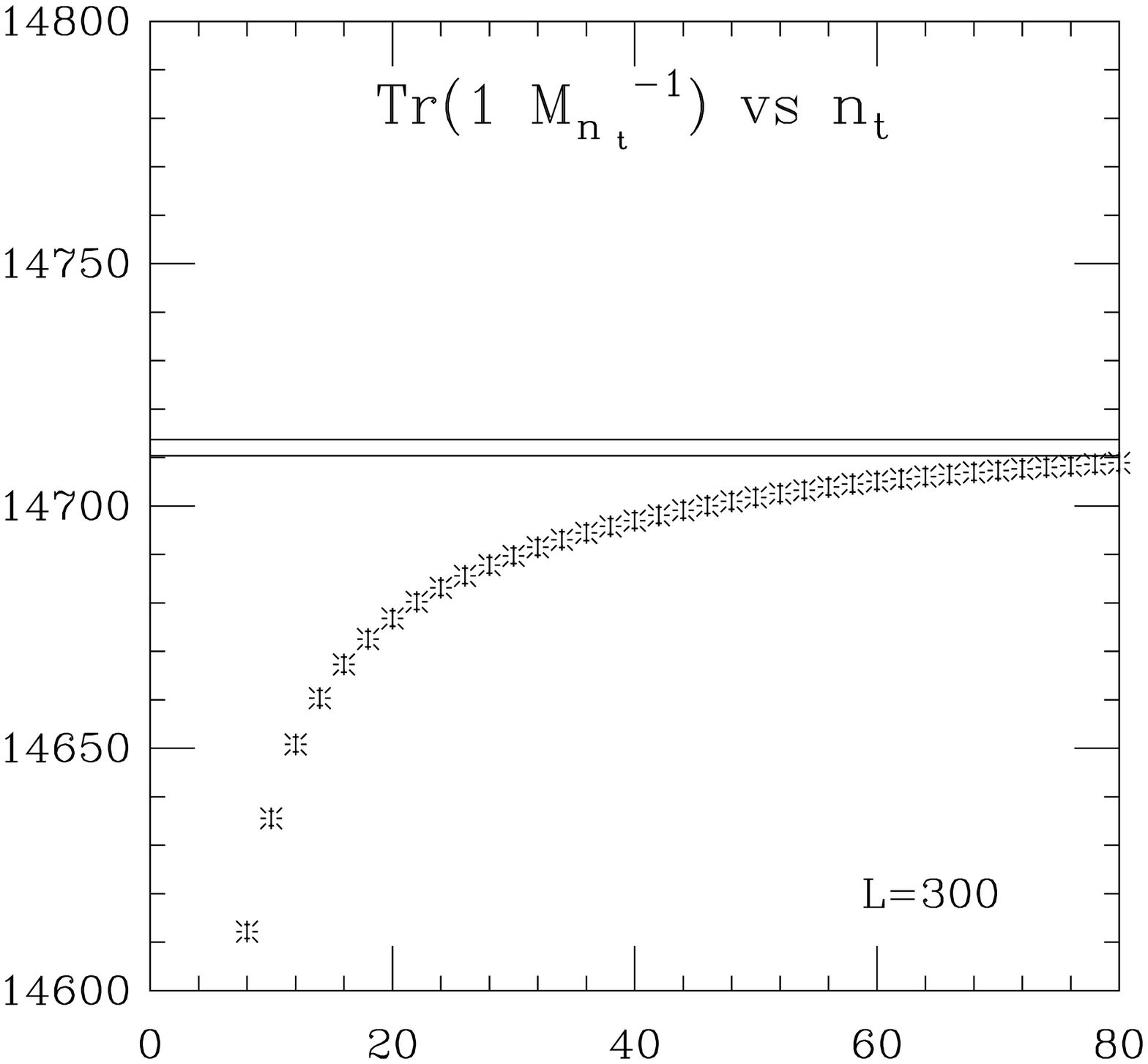}}
\caption{The disconnected loop for $\Gamma=\mathbb{1}$ as a function of the
  number of iterations used in the inverter for $\mathrm{M}^{-1}$ for
  $\kappa=0.166$. The loop is shown for (left) $L=1$ where the horizontal
  line shows the value at convergence~( $n_c=480$) and (right), with errors,
  $L=300$.}
\label{fig1:convergence}

\end{figure}

The truncated solver method~(TSM) relies on $\mathrm{Tr}(\Gamma
\mathrm{M}^{-1}_{n_t})$ coming close to the convergent value after
only a few iterations of the inverter. We found
this to be true for all $\Gamma$s studied and the example of $\Gamma=\mathbb{1}$
is shown in figure~\ref{fig1:convergence}. Clearly the trace is close
to the limiting value after 20 iterations~(compared to the 480
iterations needed for convergence). Proceeding to the calculation of
the optimal values for $n_t$ and $L_1/L_2$, we use $\mathrm{Tr}(\Gamma
\mathrm{M}^{-1}_{n_t})$ and $\mathrm{Tr}[\Gamma
(\mathrm{M}^{-1}_{n_c}-\mathrm{M}^{-1}_{n_t})]$ calculated on a single
set of stochastic estimates, $L=300$, for $n_t=2$ to $100$ in steps of
2 iterations to estimate $f_1$, $f_2$ and $f'_2$ as functions of
$n_t$. Using eqn.~\ref{optvalues} we obtain the optimal values
given in table~\ref{table:optvalues}. The results are presented for a
subset of $\Gamma$s and show that all values for $n^{opt}_t$ are small, but
also that $n^{opt}_t$ and $L_1/L_2$ depend on the $\Gamma$ used.

No error analysis has been attempted for these values and they should
be considered rough estimates. However, we have increased the number
of stochastic estimates to $500$ and no significant change in the
results was found. Using $n^{opt}_t$ and $L_1/L_2$ we can calculate
the gain in computer time using the TSM at fixed cost this time with
$L_1$ and $L_2$ independent stochastic sources. The cost, to be
inserted in eqn.~\ref{costtsm}, is set by generating $300$ stochastic
estimates of $\mathrm{Tr}(\Gamma\mathrm{M}^{-1})$. The gain corresponds to
\begin{eqnarray}
\mathrm{Gain} & = & \frac{\mathrm{Var}[\mathrm{Tr}(\Gamma \mathrm{M}^{-1})]}{\mathrm{Var}[\mathrm{Tr}(\Gamma \mathrm{M}^{-1})][\mathrm{TSM}]}
\end{eqnarray}
Table~\ref{table:optvalues} shows the TSM to result in significant
gains for all $\Gamma$s studied, including $\Gamma=\mathbb{1}$. Note that, if
time partitioning is not used in nominator and denominator these
numbers are likely to be much larger.

\begin{table}
\centerline{
\begin{tabular}{|l|l|l|l|l|l||l|l|l|l|l|}\hline
  $\mathrm{E}[\mathrm{Tr}(\Gamma\mathrm{M}^{-1})]$ &\multicolumn{5}{c||}{TSM} &\multicolumn{5}{c|}{TSM+HPE}\\\hline
$\Gamma$ & $\mathbb{1}$ & $\gamma_3$ & $\gamma_1\gamma_2$ & $\gamma_5$ & $\gamma_3\gamma_5$& $\mathbb{1}$ & $\gamma_3$ & $\gamma_1\gamma_2$ & $\gamma_5$ & $\gamma_3\gamma_5$\\\hline
$n^{opt}_t$ & 50 & 27 & 14  & 18 & 18& 66 & 78 & 50  & 78 & 90\\
$L_1/L_2$ & 23 & 21 & 32 & 28 & 30 & 26 & 25 & 21 & 26 & 26 \\
$m$ &     &    &    &    &    & 4  & 4  & 4  & 8 & 8 \\\hline
Gain &  5 & 5 & 10 & 8 & 8&  8 & 11 & 19 & 24 & 29\\\hline
\end{tabular}}
\caption{Optimal values for $n_t$ and $L_1/L_2$ for a subset of the
  $\Gamma$s studied, calculated using $L=300$ for $\kappa=0.166$. The gains
  obtained for the estimate of $\mathrm{Tr}(\Gamma\mathrm{M}^{-1})$ using
  these optimal values {\bf at fixed cost} are also shown. Where our
  method is combined with the HPE technique, $m$ indicates the order
  used.}
\label{table:optvalues}
\end{table}

\begin{figure}
\centerline{
\includegraphics[width=.4\textwidth]{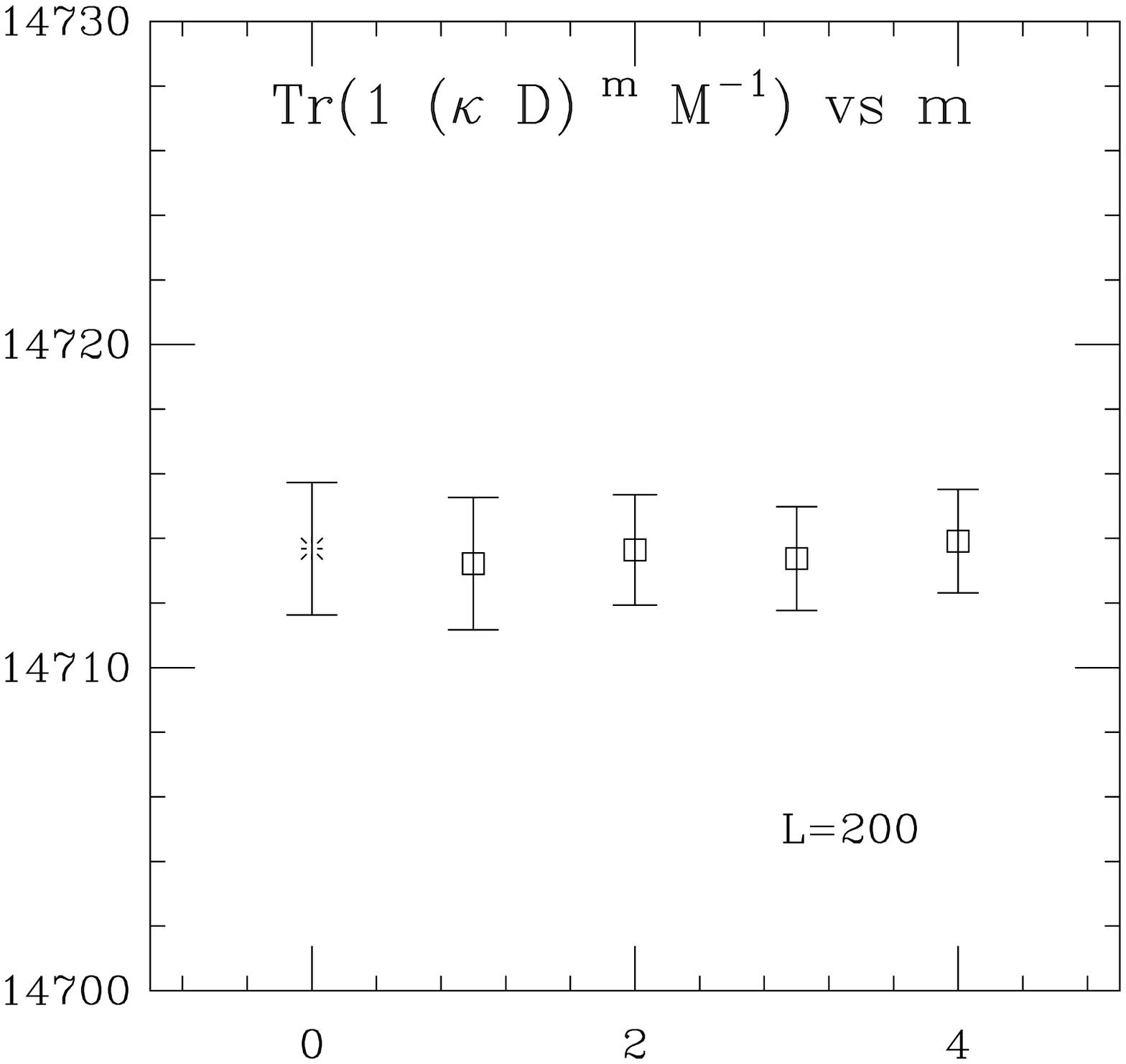}
\includegraphics[width=.38\textwidth]{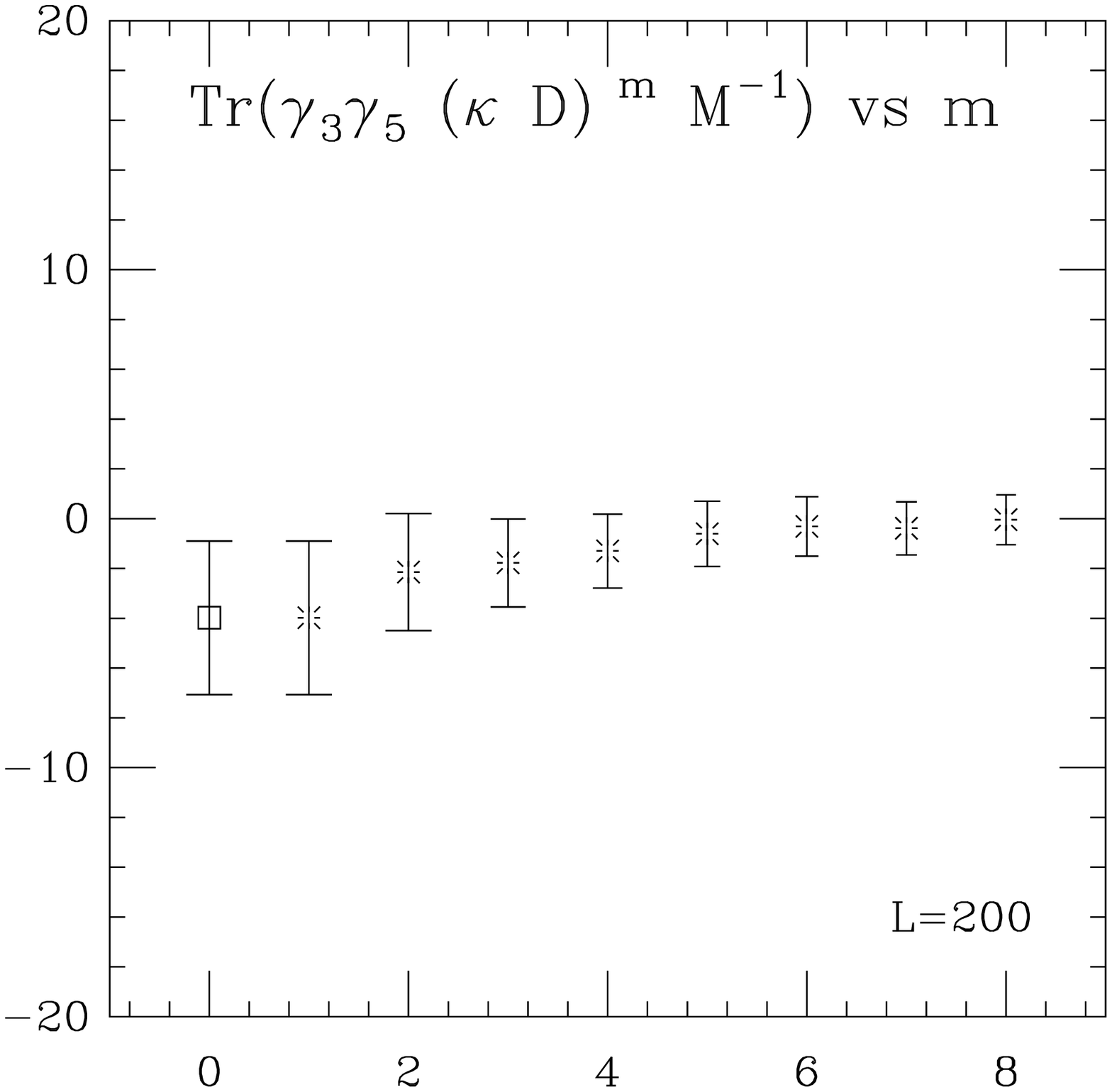}}
\caption{The disconnected loop for $\Gamma=\mathbb{1}$ and $\gamma_3\gamma_5$ as a function of the number of
  applications of $\kappa \dslash$ applied to the propagator for $\kappa=0.166$
  and $L=200$. Time partitioning has been used.
}
\label{fig2:hopex}
\end{figure}

We expect further variance reductions to be achieved when combining
our method with the HPE technique discussed in
section~\ref{newapproach}.  Figure~\ref{fig2:hopex} shows the
disconnected loop for $\kappa=0.166$, which corresponds to about
$20\%$ below the strange quark mass, for $\Gamma=\mathbb{1}$ and
$\gamma_3\gamma_5$. We see that for $\Gamma=\mathbb{1}$ the variance does
not reduce significantly as $\kappa\dslash$ is applied up to the limit
of $m=4$. This is also the case for $\Gamma=\gamma_4$. However, for
all other $\Gamma$s significant reductions in the variance are found, as
seen for $\gamma_3\gamma_5$.

Once combined with the TSM, the optimal values, $n^{opt}_t$ and
$L_1/L_2$ must be recalculated. Table~\ref{table:optvalues} shows that
$n^{opt}_t$ increases compared to using TSM alone, however, it is
still much less than $n_c=480$. With these values increased gains are
obtained for all $\Gamma$s; most notably for $\gamma_3\gamma_5$ an overall gain of a
factor of roughly $30$ is obtained. These factors were calculated
taking into account the cost of the applying the $\dslash$; for
example application of $\dslash^4$ corresponds to $5\%$ of the cost of
a propagator inversion with $n_t=66$. It may be possible to increase
the gain for $\Gamma=\mathbb{1}$, $\gamma_\mu$ and $\sigma_{\mu\nu}$ by explicitly
calculating the 4th and 6th order in the HPE.

\subsection{Effect of decreasing the quark mass}
The results presented so far have been for a quark mass slightly below
the strange quark mass. If the quark mass is reduced further,
table~\ref{table:massdep} shows that down to $m_{PS}\approx 300$~MeV there
is no significant change in the values for the TSM method. As
expected, the HPE technique becomes less effective as the quark mass
decreases and this is reflected in the drop in the factors for the
combined TSM$+$HPE.  Nevertheless, at $300$~MeV the gain is still $\geq
2$ times that for the TSM method alone for some of the $\Gamma$s.

\begin{table}
\centerline{
\begin{tabular}{|l|l|l|l|l|l||l|l|l|l|l|}\hline
  Gain & \multicolumn{5}{c||}{TSM}& \multicolumn{5}{c|}{TSM+HPE}\\\hline
  $m_{PS}$ & $\mathbb{1}$ & $\gamma_3$ & $\gamma_1\gamma_2$ & $\gamma_5$ & $\gamma_3\gamma_5$& $\mathbb{1}$ & $\gamma_3$ & $\gamma_1\gamma_2$ & $\gamma_5$ & $\gamma_3\gamma_5$\\\hline
  600~MeV  & {$5$} &5 &10 & {$8$} & 8 & {$8$}& 11 & 19 & 24 &{$29$} \\
  450~MeV  & {$5$} &5 &10&  {$8$} & 8  & {$7$}&11 & 17 & 22 &{$24$} \\
  300~MeV  & {$5$} &5 &10 & {$8$} & 8 & {$6$}& 9 & 14 & 17 & {$18$} \\\hline
\end{tabular}}
\caption{The variation in the gains for $\mathrm{Tr}(\Gamma \mathrm{M}^{-1})$ as the
  quark mass is decreased. }
\label{table:massdep}
\end{table}

\subsection{Using a different solver}
\label{bicgstab}
The results in the previous sections were obtained using the conjugate
gradiant~(CG) algorithm in the solver. We are repeating the study
using BiCGStab to see whether we can also achieve high gains with a
more optimized solver. BiCGStab converges in less iterations than CG,
for example, $n_c=156$ compared to $480$ for CG at $\kappa=0.166$.
However, each iteration is more expensive.  Furthermore, BiCGStab does
not converge smoothly. This means we cannot calculate optimal values
for $n_t$ and $L_1/L_2$~(which depend on $\partial f_2/ \partial n_t$). However, we
can fix $L_1/L_2\approx f_1/f_2$ by requiring
Var$[\mathrm{Tr}(\Gamma\mathrm{M}^{-1}_{n_t})]\approx$Var$[\mathrm{Tr}(\Gamma(\mathrm{M}^{-1}_{n_c}
- \mathrm{M}^{-1}_{n_t}))]$ and vary $n_t$ to find the best gain.
Initial results using $n_t=14$ give, for example, gains of $9$ and
$24$ for $\Gamma=\gamma_3\gamma_5$ using the TSM and TSM$+$HPE respectively, similar
to the factors obtained using the CG solver.

\section{Summary}
The truncated solver method works well, providing gains in computer
time of factors of $4-12$ for the disconnected loop, depending on the
operator, for quark masses in the range of $m_{PS}=600-300$~MeV. The
method is easy to implement, independent of the quark action and, as
we have shown, can be combined with other methods like the HPE
technique to obtain gains of factors of around $30$ for some
operators.  Future work will include combining our method with the
truncated eigenmode approach and a study of the size of the gauge
noise.

\section*{Acknowledgments}
S.~Collins acknowledges financial support from the
Claussen-Simon-Foundation (Stifterverband f\"ur die Deutsche
Wissenschaft). This work has also been supported by the EC Hadron
Physics I3 Contract RII3-CT-2004-506087, the BMBF Project 06RY258 and
the DFG. We thank Mike Clark, Chris Michael and Hartmut Neff for
discussions.

\end{document}